\documentclass[12pt]{iopart}
\usepackage{graphicx}
\begin{document}

\title[Arc temperatures \& iterative OES]{Arc temperatures in a circuit breaker experiment from iterative analysis of emission spectra}

\author{Steffen Franke$^1$, Ralf Methling$^1$, Dirk Uhrlandt$^1$, Sergey Gortschakow$^1$, Frank Reichert$^2$, and Arkadz Petchanka$^2$}

\address{$^1$Leibniz Institute for Plasma Science and Technology, Felix-Hausdorff-Str. 2, 17489 Greifswald, Germany\\
$^2$ Siemens Gas and Power GmbH \& Co. KG, Nonnendammallee 104, 13629 Berlin, Germany}
\ead{steffen.franke@inp-greifswald.de}

\vspace{10pt}
\begin{indented}
\item[]April 2020
\end{indented}

\begin{abstract}
A switching-off process very similar to those in real high-voltage self-blast circuit-breakers is emulated in a model chamber to study the arc properties by optical emission spectroscopy. The arc is operated in a chamber filled with SF$_6$ between a pin-tulip contact system enclosed by a PTFE nozzle. Transparent windows in the chamber wall and a slit in the nozzle enable an optical investigation of the arc cross section less than one millisecond before current zero. The side-on radiance of a fluorine atom lines has been measured with an imaging spectroscopic system. Considering rotational symmetry of the arc the corresponding radial emission coefficients have been determined by Abel inversion. The radial temperature profiles are calculated from the emission coefficients including pressure measurements and the corresponding plasma composition. In contrast to previous studies, a specific iterative approach is proposed to consider the optical thickness of the plasma. A maximum optical thickness between 0.2 and 0.4 has been observed. Temperature profiles corrected for absorption effects show increasing maximum temperature in the arc axis with decreasing current towards current zero. The effect of the plasma composition on absolute arc temperatures is discussed. The temperature profiles can be used to validate CFD simulations of the switching-off process in the model chamber.
\end{abstract}

%

%
%
%

\section{Introduction}

High-voltage self-blast circuit-breakers are still of particular interest, although a number of replacement gases for SF$_6$ are under consideration. However, the exceptional performance of SF$_6$ is required at the limits of switching applications at high voltages and high currents. 

Detailed modelling of the processes in the arc chamber and the forecast of the breaking capability are availably for around 20 years nowadays  (see e.g. \cite{God00}). The simulations based on the methods of computational fluid dynamics (CFD) cover the arc plasma, the nozzle ablation and the flow behaviour at high pressures even in complex geometries. Despite of the increased accuracy of such models and of the required atomic data, validations by experimental measurements are mandatory. The validation of calculated plasma properties of the switching arc, particularly of plasma temperature profiles,  is one of the most sensitive methods to verify CFD simulations of the arc. 

Optical emission spectroscopy (OES) of the arc radiation offers a variety of methods for the spatially resolved determination of plasma temperatures and species densities, and  it is a non-invasive method \cite{Wie91}. The arc radiation is a key issue for the nozzle ablation, the pressure and temperature increase and finally the breaking capability in a self-blast circuit breaker \cite{Ruch86}. For these reasons the investigation of arc radiation by OES and comparison with radiation simulation is of particular interest.

The first challenge of OES is the requirement of an optical access to the arc in a circuit breaker chamber. This is the reason why most of OES measurements have been applied to simplified arc experiments which have been designed to emulate the typical arc behaviour in a real switching device. The second challenge is the choice of an appropriate method for spectral analysis connected with the choice of spectral lines considering their broadening and optical thickness as well as the plasma composition. Examples of previous studies with respect to these two topics are given in the following.

One option to investigate ablation controlled arcs is to place a cylindrical polytetrafluoroethylene (PTFE) nozzle between two electrodes in ambient air\cite{Koz07}. Although the geometry is strongly simplified, the general arc behaviour of a self-blast circuit breaker can be studied, because of the strong nozzle ablation and pressure increase at elevated arc currents. Arcs in a current range from 10 to 22\,kA have been investigated in \cite{Koz07} by imaging OES over the arc cross section very close to the nozzle exit where their behaviour is dominated by the PTFE gas at high pressure and flow velocity. Absolute intensities of carbon ion lines at 426 and at 658\,nm have been used for the determination of radial temperature profiles. In addition, it was shown that standard analysis of fluorine atom lines at 641 and 634\,nm results in lower temperatures because of the considerable self-absorption of these lines at pressures around 10\,bar.

Only few studies deal with experiments in more realistic chamber geometries of circuit-breakers. A flat-type gas-blast quenching chamber filled with SF$_6$ has been considered in \cite{Tan97}. Optical fibre bundles at five positions in the chamber wall have been used to couple out the arc radiation along a considerable part of the arc length. Low iron vapour concentration from electrode erosion yields iron atom lines of low optical thickness. Two lines at 426 and 442\,nm have been used in \cite{Tan97} for the determination of the arc temperature and the iron concentration at the five axial arc positions at different times before current zero and in the post-arc channel. An arc plasma homogeneous over the arc cross section has been assumed for simplicity.

A combination of OES of the entire arc and fast imaging with spectral filters has been applied to a model circuit breakers operated in SF$_6$ by \cite{Hart98}. Setups for a puffer breaker and a self-expansion breaker with optical access through two windows and strongly simplified chamber geometries have been considered. Line intensities of fluorine atom lines in the range between 623 and 642\,nm have been used for the temperature determination. Griem's derivation of Stark broadening of the lines has been applied to determine electron densities. The impact of radiation absorption of the fluorine lines on the measured line intensities has been corrected by estimations of the arc size from the fast images. Again, constant temperature over the arc cross section has been assumed for simplicity.

However, the determination of radial profiles of plasma temperatures generally requires the radiation analysis spatially resolved over the arc cross section. An experiment with optical access to the heating channel in a simplified model circuit breaker has been studied by OES in \cite{Eich12}. The chamber was filled with CO$_2$ as an attractive replacement option for SF$_6$. Nevertheless, the arc in the high-current range was dominated by ablation of the PTFE nozzle. Measurements of the absolute line intensity of the carbon ion line at 658\,nm have been used for the determination of the radial temperature profiles. Beside pressure measurements with probes, the Stark width of the carbon ion line has been compared with the measured profile width to deduce radiator density and pressure under assumption of pure PTFE gas for considered current instants. 

In contrast to the previous studies mentioned above, experiments in a SF$_6$ self-blast model circuit-breaker with a realistic inner chamber, PTFE nozzle and electrode geometry are considered in this paper. The nozzle is only modified by introducing a thin slit used for imaging OES over the arc cross section. The study is focussed on the decaying phase (less than 1\,ms before current zero crossing) of a high-current, ablation dominated arc in a model circuit breaker. Emphasis is put on the determination of the temperature profile over a large extent of the arc cross section. Therefore, line radiation of fluorine atoms is evaluated instead of ion radiation. The impact of self-absorption is considered by a specific iterative procedure, in contrast to a previous study where optically thin conditions were assumed\cite{Methling2015p163}. As in all previous studies mentioned above, local thermodynamic equilibrium in the arc plasma is required for the radiation analysis.

The paper is organized as follows: The experimental setup and parameters are given in section 2. The methods for the analysis of OES measurements are explained in section 3. Results including error estimation are given in section 4, and section 5 gives a summary and an outlook.

\section{Experiment}

The experimental setup consists of a model breaker with electrodes, nozzle and arc chamber very similar to real devices. The electrodes (pin and tulip) are made of W-Cu with pin diameter of 15\,mm. The electrodes are separated inside a PTFE nozzle with a narrow part of an inner diameter of 18\,mm followed by a larger turbulence volume of 50\,mm inner diameter surrounding the tulip. A slit of 2\,mm width, perpendicular to the nozzle axis, is used for optical investigations, which is a compromise between providing an optical access to the arc and avoiding a perturbation of the arc. The set-up is confined in a closed vessel filled with SF$_6$ at a pressure of 6\,bar (absolute). Glass windows at opposite sides of the vessel provide the optical access to the nozzle slit. The transparency of these windows has been controlled by transmission measurements after every third arc experiment. It turned out that a gas exchange including a complete exhaust of the fume is appropriate for a sufficient transmission of the windows and the vessel volume. By this procedure, the transmission was ensured to stay around 0.87. Measured intensities were corrected by this value. The discharges were operated with a sine-like current over one half-wave of 50\,Hz frequency and maximum currents of about 14\,kA (effective current 10\,kA) by means of a synthetic test circuit. A pressure probe is fixed inside the PTFE nozzle close to the arc in the vicinity of the slit and is utilized to determine the time-dependent static pressure.
\begin{figure}[ht]
	\centering
	\includegraphics[width=8cm]{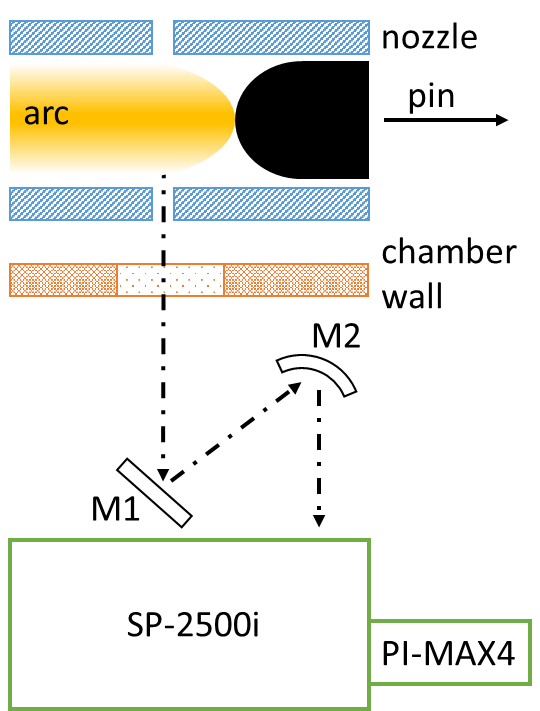}
	\caption{Scheme of the experimental setup consisting of the nozzle with slits, model breaker chamber with window, deflecting mirror (M1), focussing mirror (M2), spectrograph (SP-2500) and intensified CCD (PI-MAX4).}
	\label{fig:ExpSetup}
\end{figure} 
 
\begin{figure}[ht]
	\centering
	\includegraphics[width=16cm]{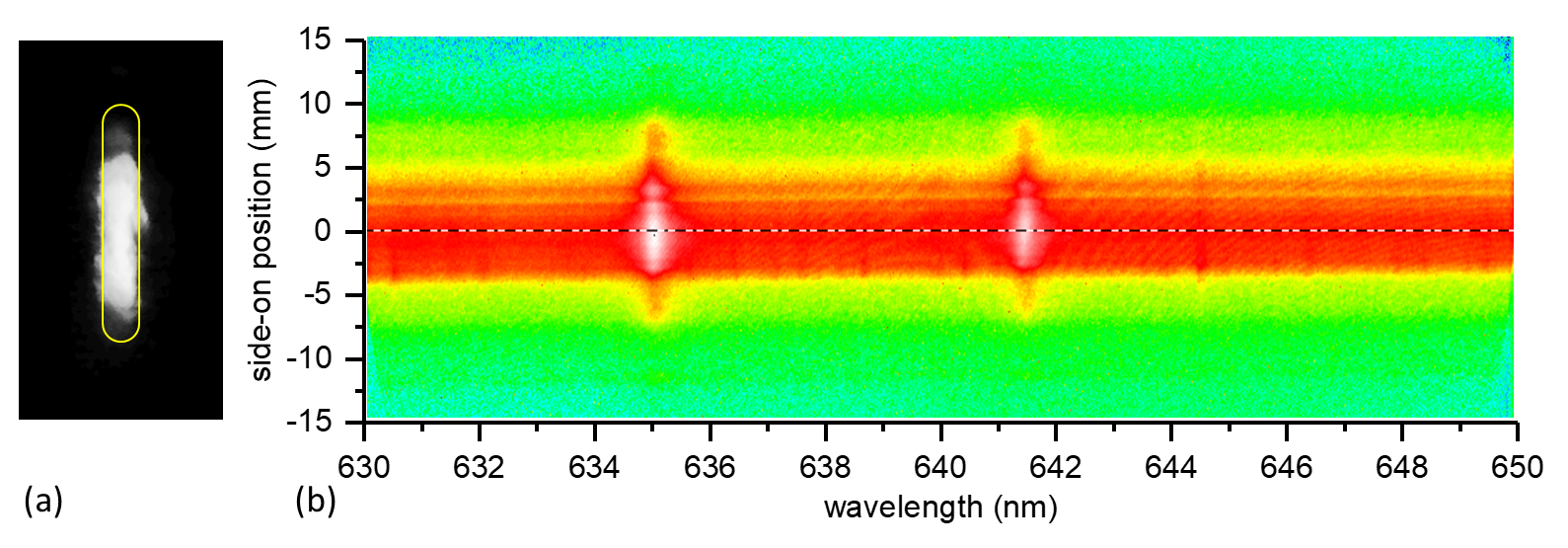}
	\caption{(a) High-speed image of the arc observed through the slit in the model breaker. The yellow line indicates the slit boundary. (b) ICCD-Image with wavelength in horizontal direction and radial side-on-position in the vertical direction.}
	\label{fig:Images}
\end{figure}

The radiating arc is imaged on the entrance slit of a Czerny-Turner spectrograph (Acton Research Corporation SpectraPro 2500i) by a spherical mirror with 0.5\,m focal length through the slit in the PTFE nozzle and the window in the vessel as illustrated in Fig.\,\ref{fig:ExpSetup}. 
With an object distance of about 3\,m a magnification of around 0.2 is achieved to record the maximum arc width by the detector. The spectrograph is equipped with an intensified CCD camera (Princeton Instruments PI-MAX2, 1024x256 pixels, 26\,$\mu$m\,x\,26\,$\mu$m) as detector, in order to obtain single 2D spectra typically with a wavelength resolution of 0.06\,nm (FWHM of the instrumental profile) for a spectral window of about 20\,nm (cf. Fig.\,\ref{fig:Images}(b)). Absolute calibration of spectra in units of spectral radiance is performed by a tungsten strip lamp (OSRAM Wi 17/G). In addition the general arc dynamics is recorded by a high-speed camera (IDT MotionPro Y4). An example of a typical high-speed image of the arc is shown in Fig.\,\ref{fig:Images}(a) taken with a frame rate of 10\,000 frames per second and an exposure time of 1\,$\mu$s.

\section{Analysis of radiation measurements}

In this section the procedures for analysing the arc radiation are described. It is expected that the arc radiation in the model breaker is dominated by atomic and ionic line radiation of the species F, C and S, because the chamber is filled with SF$_6$ and the ablation of the PTFE nozzle could be considerable. Depending on the ablation the radiation contribution of C and S at different times during the current half wave can hardly be predicted. This is one reason for choosing a atomic fluorine lines for the analysis in contrast to former works \cite{Koz07,Eich12} where carbon ion lines were considered. Another reason is that sufficient radiation from a fluorine atom line is expected also at outer radial positions of the arc with lower temperature, because of the relatively high population of excited fluorine atomic states. This makes the determination of the temperature profile over a larger arc cross section possible. 

Finally, the F\,I lines at 641\,nm and 635\,nm have been chosen because the lines are well isolated from other spectral lines and atomic data are available with sufficient accuracy. The analysis of the absolute radiation intensity of one spectral line is considered because application of the two-line method would require the simultaneous measurement of neighbouring lines with energy levels of the radiating states which differ appropriately. In contrast to that, recording of the radiance of one line is possible with high wavelength resolution and applying only one measurement.
Disadvantages of the F\,I lines are the higher optical thickness as a result of the higher population of the lower state and the lower energy of the upper state in comparison with ion lines, in particular carbon ion lines used in \cite{Koz07,Eich12}. The latter leads to an increase of the uncertainty in the temperature determination with respect to the measurement error in the emission coefficient -- if optical thickness is not accounted for. 

\begin{figure}[ht]
	\centering
	\includegraphics[width=8cm]{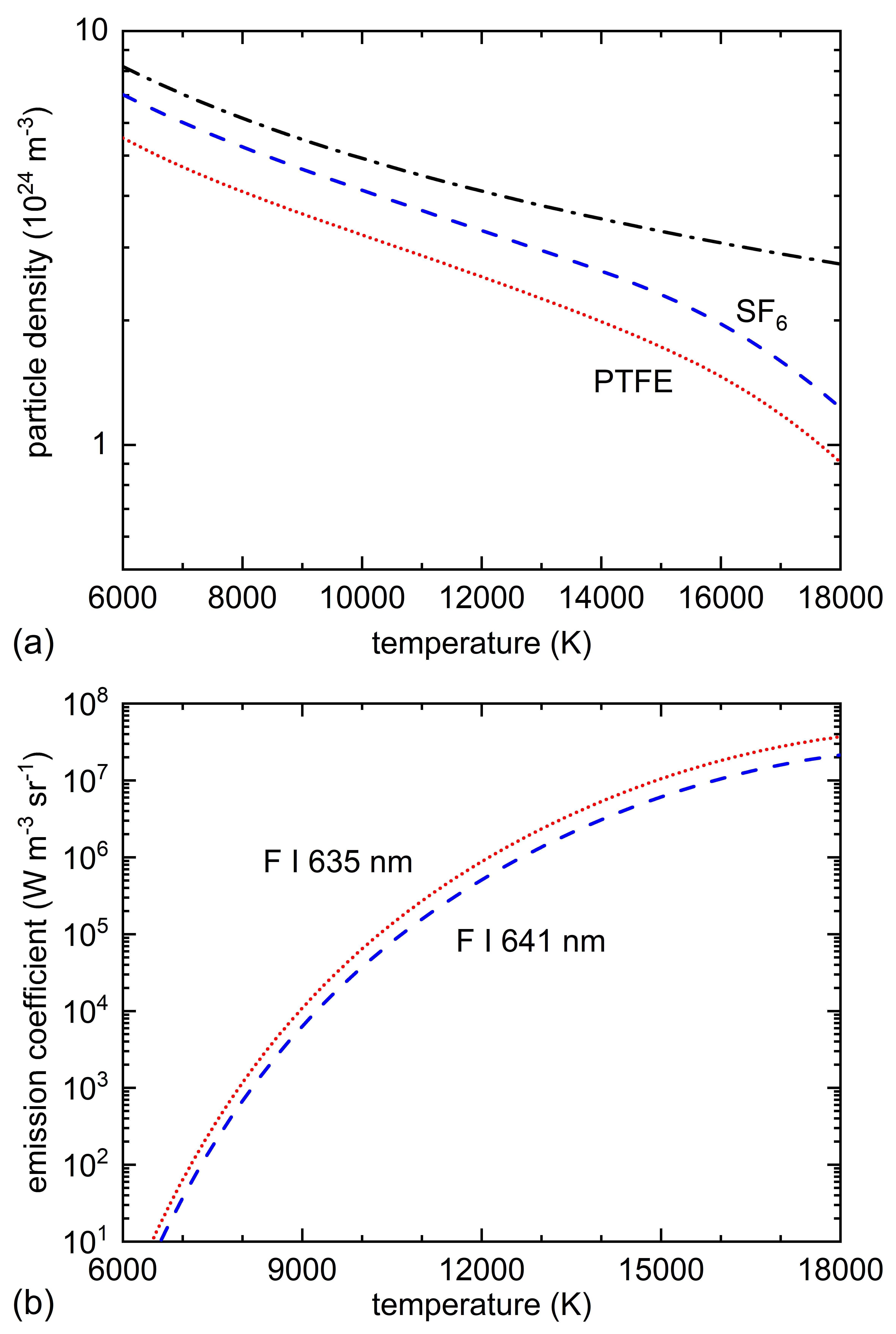}
	\caption{(a) Calculated atomic fluorine particle density for pure SF$_6$ gas (dashed) and for pure PTFE gas (dotted) at a total pressure of 6.8\,bar. Dash-dotted line represents the total particle density following ideal gas law. (b) Emission coefficient of the spectral line of F\,I at 641\,nm (dashed) and of F\,I at 634\,nm (short dashed) for pure SF$_6$ plasma.}
	\label{fig:dens-eps-theo}
\end{figure} 
The plasma composition for the inner part of the nozzle depending on temperature $T$ and pressure $p$ is calculated in advance assuming local thermodynamic equilibrium according to standard procedures. Two limiting cases are considered: the composition for pure vaporized PTFE and for pure filling gas SF$_6$ for pressures given from the pressure measurements. Figure\,\ref{fig:dens-eps-theo}(a) presents atomic fluorine particle density for pure SF$_6$ gas at a total pressure of 6.8\,bar as well as for pure PTFE gas. For comparison the total particle density following from ideal gas law is drawn as dash-dotted line. It can be seen that the plasma composition is dominated by fluorine and in the given temperature range almost following the ideal gas law, except for temperatures above 15\,000\,K where atomic fluorine density starts to deviate due to significant ionization. Making use of the fluorine atom densities $n_g(T,p)$, the emission coefficient $\varepsilon$ of the F\,I lines is calculated according to
\begin{equation}\label{eqn:epsTp}
  \varepsilon(T,p) = \frac{hc}{4\pi \lambda_0} A_{ul} \frac{g_u n_g(T,p)}{Z(T)} 
  \mathrm{exp}\left(-\frac{E_u}{kT}\right) 
\end{equation}
assuming Boltzmann distribution of excited states and using transition probability $A_{ul}$ and upper level energy $E_u$ from NIST data base \cite{NIST2020}. Here $Z(T)$ is the partition function which is calculated according to the Planck-Larkin relation\cite{Ebeling1973a, Kahlbaum1990p753}. $g_u$ is the statistical weight, $c$ the speed of light in vacuum, $k$ and $h$ are the Boltzmann and Planck constants, respectively. Atomic data for both transitions of atomic fluorine are given in Tab.\,\ref{tab:atomicData}. 
\begin{table}
	\centering
	\caption{Atomic data for two transitions of atomic fluorine according to NIST\cite{NIST2020} database. $\lambda_0$, $A_{ul}$, $E_u$, and $g_u$ are the centre wavelength, transition probability, level energy, and statistical weight, respectively. Indices $u$ and $l$ refer to the upper and lower level.}
	\begin{tabular}{llllll}
		\hline
		$\lambda_0$ (nm) & A$_{ul}$ (1/s) & E$_l$ (eV) & g$_l$ & Eu (eV) & g$_u$ \\ \hline
		634.851 & $2.32\cdot10^7$ & 12.730752 & 4 & 14.683178 & 4 \\ 
		641.365 & $1.36\cdot10^7$ & 12.750582 & 2 & 14.683178 & 4 \\ \hline
	\end{tabular}
	\label{tab:atomicData}
\end{table}
The emission coefficient of the spectral line F\,I at 641\,nm as a function of temperature for pure SF$_6$ gas and a total pressure of 6.8\,bar is shown in Fig.\,\ref{fig:dens-eps-theo}(b) as well as for F\,I at 635\,nm. A consequence of the strong gradient in the emission coefficient as a function of temperature is the comparably small experimental uncertainty in temperature determination. Even if the emission coefficient is just known with an uncertainty of about 30\, $\%$, the uncertainty in the temperature will be only around 3\,$\%$.

\begin{figure}[ht]
	\centering
	\includegraphics[width=7cm]{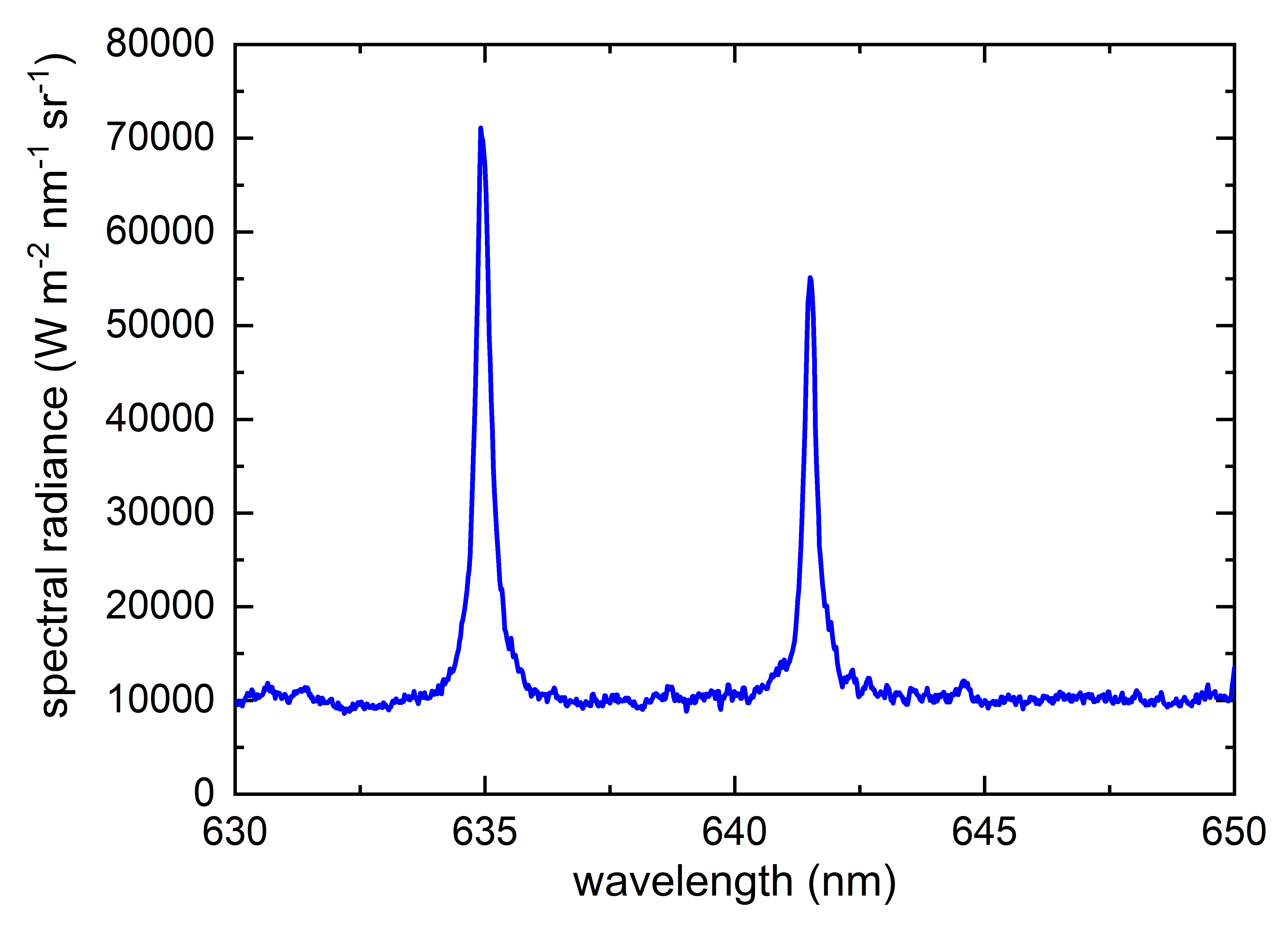}
	\caption{Measured and calibrated spectral radiance in the range from 630 to 650\,nm for the side-on position $y_0$ corresponding to the centre of the arc, recorded 1.25\,ms before current zero at 3.3\,kA instantaneous current.}
	\label{fig:Radiance}
\end{figure}

The spectral radiance $L_\lambda(y,\lambda)$ measured in the wavelength range $\lambda = 630$ to 650\,nm and for a range of side-on positions $y$ of about 18\,mm is, in a first step, calibrated with respect to intensity by a tungsten strip lamp (OSRAM Wi 17/G) and with respect to wavelength by a Hg/Ar penray (LOT-Oriel LSP035). An example spectrum is shown in Fig.\,\ref{fig:Radiance} (see also Fig.\,\ref{fig:Images}). 

In a second step, the line radiance $L(y)$ of the F\,I line at 641\,nm is determined for every side-on position $y$ by appropriate integration of the spectral radiance. If the spectral lines are well separated, this can be done by simple numerical integration. A constant offset is subtracted in advance to account for a continuum contribution to the emission. 

\begin{figure}[ht]
	\centering
	\includegraphics[width=8cm]{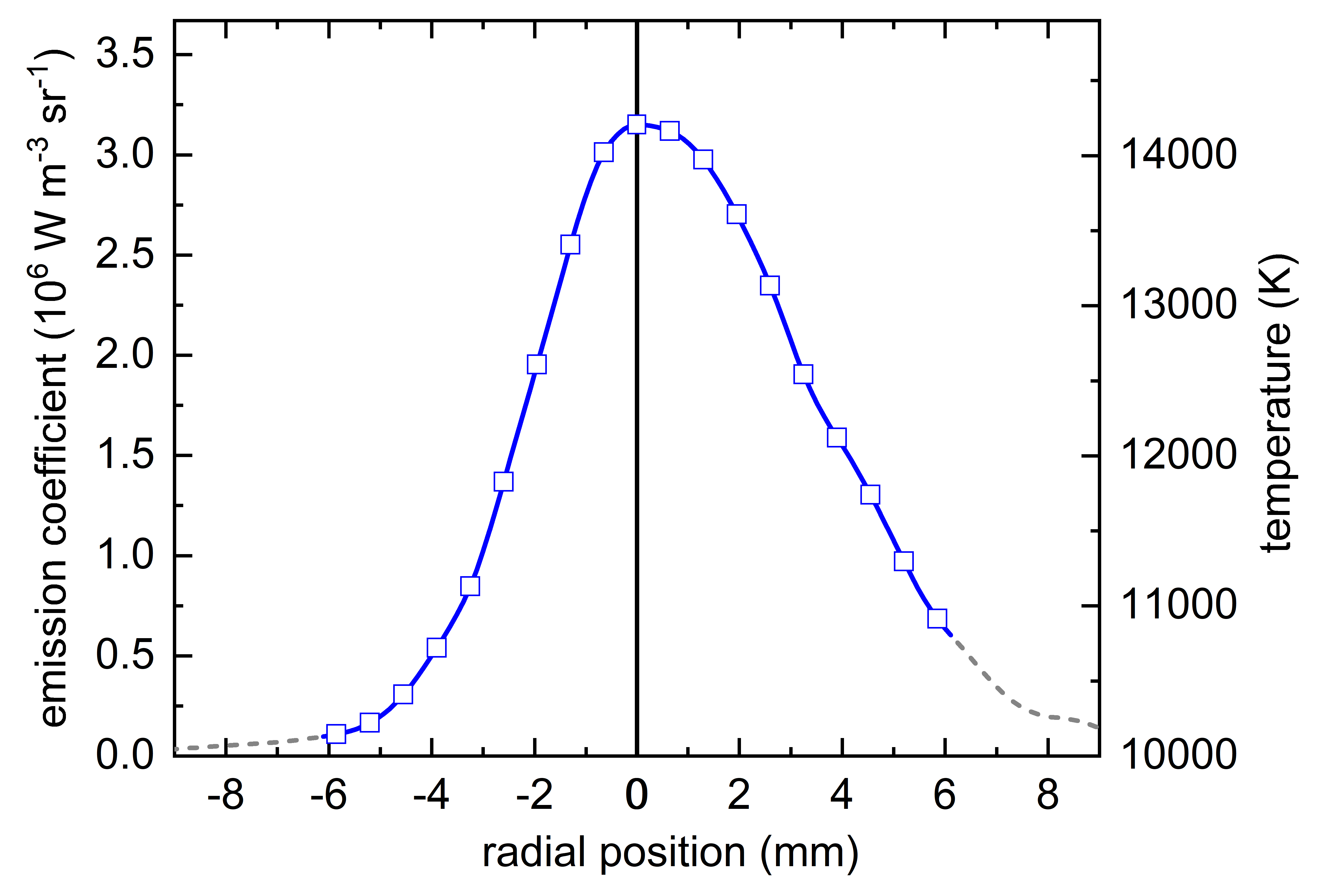}
	\caption{Emission coefficient for F\,I at 641\,nm after Abel inversion (left) and radial temperature profile from pre-calculated emission coefficients (right), 1.25\,ms before current zero at 3.3\,kA instantaneous current.}
	\label{fig:Eps-T-r}
\end{figure}

According to the standard procedure for the analysis of a spectral line in a plasma of low optical thickness, the line radiance $L(y)$ is considered as the accumulation of the emission along a line of sight through the arc cross section, and the radiation absorption along the line of sight is neglected\cite{Lochte-Holtgreven1995a}. Then, assuming in addition rotational symmetry of the arc, the line radiance $L(y)$ is connected to the emission coefficient $\varepsilon(r)$ as a function of the radial position $r$ via direct Abel transformation over the radius of the nozzle volume $R$:
\begin{eqnarray}\label{eqn:Abel1}
  L(y) &= &2\int\limits_y^R \frac{\varepsilon(r) r {\rm d}r}{\sqrt{r^2-y^2}}
  \equiv\hat{\bf A}\varepsilon(r)\\ \label{eqn:Abel2}
  \varepsilon(r) &= & -\frac{1}{\pi}\int\limits_r^R \frac{{\rm d}L(y) }
  {{\rm d}y} \frac{{\rm d}y}{\sqrt{y^2-r^2}} \equiv \hat{\bf A}^{-1}L(y).
\end{eqnarray}
Hence, the radially dependent emission coefficient is obtained from inverse Abel transformation (Abel inversion) $\hat{\bf A}^{-1}$ of the line radiance  $L(y)$. Finally, comparison of the experimentally determined values of   $\varepsilon(r)$ according to Eqn.\,(\ref{eqn:Abel2}) with the pre-calculated emission coefficients $\varepsilon(T,p)$ according to Eqn.\,(\ref{eqn:epsTp}) for each radial position $r$ and for the measured pressure $p$ yields the radial temperature profile $T(r)$. This method is called calibrated line method or single line method, because it requires an absolute calibration of radiances and can be performed on a single spectral line (as long as the radiator density is known, e.g. from plasma composition calculation). Figure\,\ref{fig:Eps-T-r} presents the emission coefficient for F\,I at 641\,nm after Abel inversion, and the according radial temperature profile from pre-calculated emission coefficients for the measurement given in Fig.\,\ref{fig:Radiance}. Temperatures are evaluated up to a radial position of 6--7\,mm, because uncertainty in the emission coefficient increases significantly at outer radial positions (indicated by dashed lines).

\subsection{Consideration of absorption effects}

The question now is, how can one be sure that the plasma is optically thin? From visual inspection of the spectral line an optically thick plasma can be identified, if a self-reversal of the spectral line contour occurs which may happen in optically thick inhomogeneous plasmas\cite{Zwicker1995a}. If the plasma is spatially homogeneous the line centre will flatten when reaching the Planck limit\cite{Franke2014a}. If non of both effects happens, further options are to investigate the optical depth by absorption measurements using an additional light source placed behind the  plasma or using the back-reflection of plasma radiation itself by placing a focussing mirror behind the arc, which of course requires some experimental effort.

\subsubsection{Optical depth estimation from effective values.}

However, there is a simple method to estimate optical thickness of a plasma from measurements of side-on-intensities. The main point is, that the absorption coefficient $\kappa$ is related to the spectral emission coefficient $\varepsilon_\lambda$ by the Planck function $B_\lambda$ given in units of spectral radiance via Kirchhoff's law 
\begin{equation}\label{eqn:Kirchhoff}
 B_\lambda(\lambda,T) = \frac{\varepsilon_\lambda(\lambda)}{\kappa(\lambda)}~.
\end{equation}
That means, knowing the emission coefficient and the temperature it is be possible to determine the absorption coefficient and thus the optical depth. Assuming a Lorentz line profile
\begin{equation}\label{eqn:Lorentz}
  P_\lambda = \frac{2}{\pi \Delta\lambda} \frac{1}{1+\left(\frac{\lambda-\lambda_0}{\Delta\lambda/2}\right)^2}
\end{equation}
with centre wavelength $\lambda_0$ and full widht at half maximum $\Delta\lambda$ the spectral  emission coefficient in the centre of the spectral line $\varepsilon_{\lambda,0}$ can be calculated from the measured and line integrated emission coefficient $\varepsilon$  by
\begin{equation}\label{eqn:eps0}
  \varepsilon_{\lambda,0} =  \varepsilon P_{\lambda,0} = \varepsilon   \frac{2}{\pi\Delta\lambda} 
\end{equation}
and hence for an optical path length $l$ the optical depth in the line centre $\tau_0$ is obtained from
\begin{equation}\label{eqn:tau0}
  \tau_0 = \kappa_0 l = \varepsilon_{\lambda,0} \frac{l}{B_\lambda} = \varepsilon  \frac{2}{\pi\Delta\lambda} \frac{l}{B_\lambda}
\end{equation}
From evaluation of a spectral line by calibrated line method (see above) most of the data already are known. The spectrally integrated emission coefficient $\varepsilon$ is determined according to calibrated line method in this example to be $3.2\cdot10^6\,\mathrm{W m^{-3} sr^{-1}}$ in the centre of the arc. With an axis temperature of around 14\,200\,K the Planck function value is $2.9\cdot10^5\,\mathrm{W m^{-2} nm^{-1} sr^{-1}}$ at 641\,nm. The optical path is not longer than 18\,mm. From Fig.\,\ref{fig:Radiance} a full width at half maximum $\Delta\lambda$ of 0.3\,nm can be estimated - neglecting any instrumental profiles. Altogether an optical depth of around 0.43 is obtained.  This value indicates that data evaluation should be carefully checked for any absorption effects. However, this is an upper limit and the estimation could be refined by using the average emission coefficient along the line of sight and not the maximum axis value. The average emission coefficient, obtained from
\begin{equation}\label{eqn:tau0}
\left<\varepsilon\right> = \int_{0}^{R} \varepsilon(r) \mathrm{d} r /R~,
\end{equation}
is in this case about $0.91\cdot10^6\,\mathrm{W m^{-3} sr^{-1}}$ which yields an optical depth of 0.12. This value might give a lower limit for the optical depth in the line centre.

\subsubsection{Optical depth estimation from line of sight integration in line centre.}

The approach could be further refined by calculating the local absorption coefficient in the line centre $\kappa_0$ from the spectral emission coefficient $\varepsilon_{\lambda,0}$ and integrating along the line of sight through the arc axis.
\begin{equation}\label{eqn:tau0integrated}
  \tau_0 = 2\int\limits_0^R \frac{\varepsilon_{\lambda,0}(r)}{B_{\lambda,0}(T(r))}
  \mathrm{d}r
\end{equation}
Knowing the radial temperature profile from the calibrated line evaluation above, Planck function $B_{\lambda,0}(T(r))$ can be determined, and after Abel inversion of the side-on radiance in the line profile centre $L_{\lambda,0}$ the local spectral emission coefficient $\varepsilon_{\lambda,0}$ in the line centre is obtained. For the data presented above an optical depth of 0.15 is calculated here for the line of sight through the arc centre, which is in-between those two values calculated before and give still rise to a deeper analysis of absorption effects. But is there any chance not only to estimate optical depth but also to correct the experimental values for absorption effects?
  
\subsubsection{Optical depth correction by an iterative procedure.}

\begin{figure}[ht]
	\centering
	\includegraphics[width=12cm]{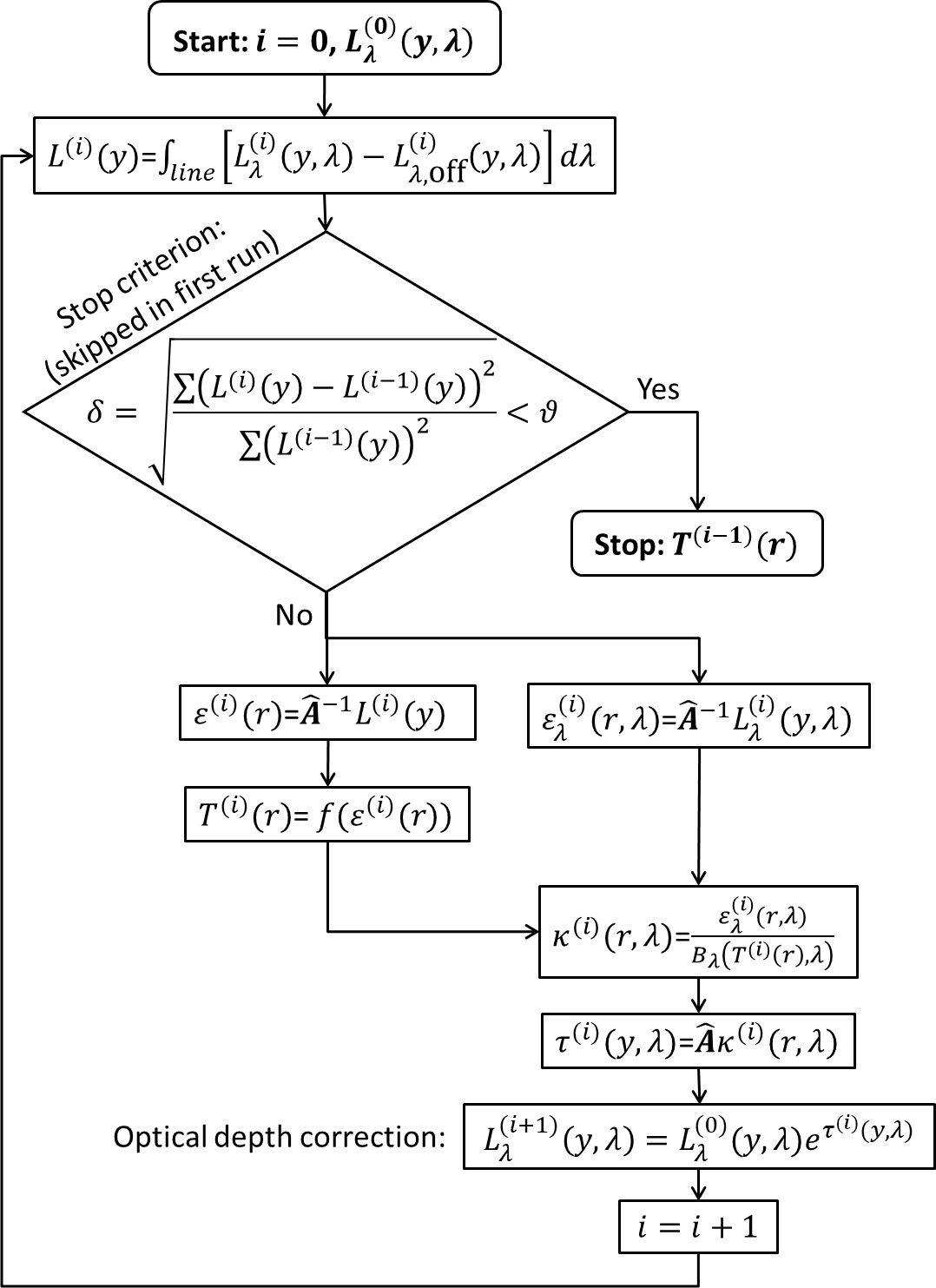}
	\caption{Iterative scheme for calibrated line method taking account of absorption.}
	\label{fig:Schema}
\end{figure} 
In fact, the measured side-on radiance in the line centre $L_{\lambda,0}$ can be corrected with respect to the optical depth $\tau_0$ calculated from Eqn.\,(\ref{eqn:tau0integrated}) by inverting the Lambert-Beer law:
\begin{equation}\label{eqn:odcorrection}
L_{\lambda,0}^{\mathrm{corr}} = L_{\lambda,0}{\rm exp}\left(+\tau_0\right).
\end{equation}
$L_{\lambda,0}^{\mathrm{corr}}$ represents a first estimate of the line radiance increased according to the radiation absorbed along the line of sight at the line centre. This procedure can be repeated iteratively. The full iterative scheme taking into account absorption at different wavelengths of the line profile and at different side-on positions is given in Fig.\,\ref{fig:Schema}. The scheme is started with the measured side-on radiance $L_{\lambda}^{(0)}$ at all side-on positions $y$ and all wavelengths $\lambda$. The top left branch of the scheme maps the calibrated line method described above (skipping the stop criterion in the first run). Notice, that a continuum contribution $L_{\lambda,\mathrm{off}}(y)$ to the side-on spectral radiance $L_\lambda(y)$ is subtracted from the line profile before spectral integration. Abel inversion then delivers the line emission coefficient $\varepsilon$. From pre-calculated emission coefficients the temperature profile $T(r)$ is obtained. This step is designated by the function $f(\varepsilon(r))$ in the scheme. The bottom right branch of the scheme starts with an Abel inversion of the side-on radiance $L_{\lambda}(y,\lambda)$ at all wavelengths of the line profile. Making use of the temperature profile $T(r)$ obtained from the left branch the absorption coefficient $\kappa(r,\lambda)$ is calculated for all radial positions and wavelengths. After direct Abel transformation applied to the absorption coefficient the optical depth $\tau(y,\lambda)$ for all side-on positions and wavelengths is obtained. Now, the side-on spectral radiance is corrected with respect to the obtained optical depth. Finally, the counter is increased by one and the scheme is iterated. The stop criterion $\delta$ measures the root mean square of the present side-on line radiance to the previous side-on line radiance and stops iteration, if it falls below the limiting value $\vartheta$, e.g. $10^{-3}$. For the cases considered in this study not more than 10 iterations have been performed to achieve convergence. 

\section{Results and discussion}

\subsection{Iterative optical emission spectroscopy}

\begin{figure}[ht]
	\centering
	\includegraphics[width=8cm]{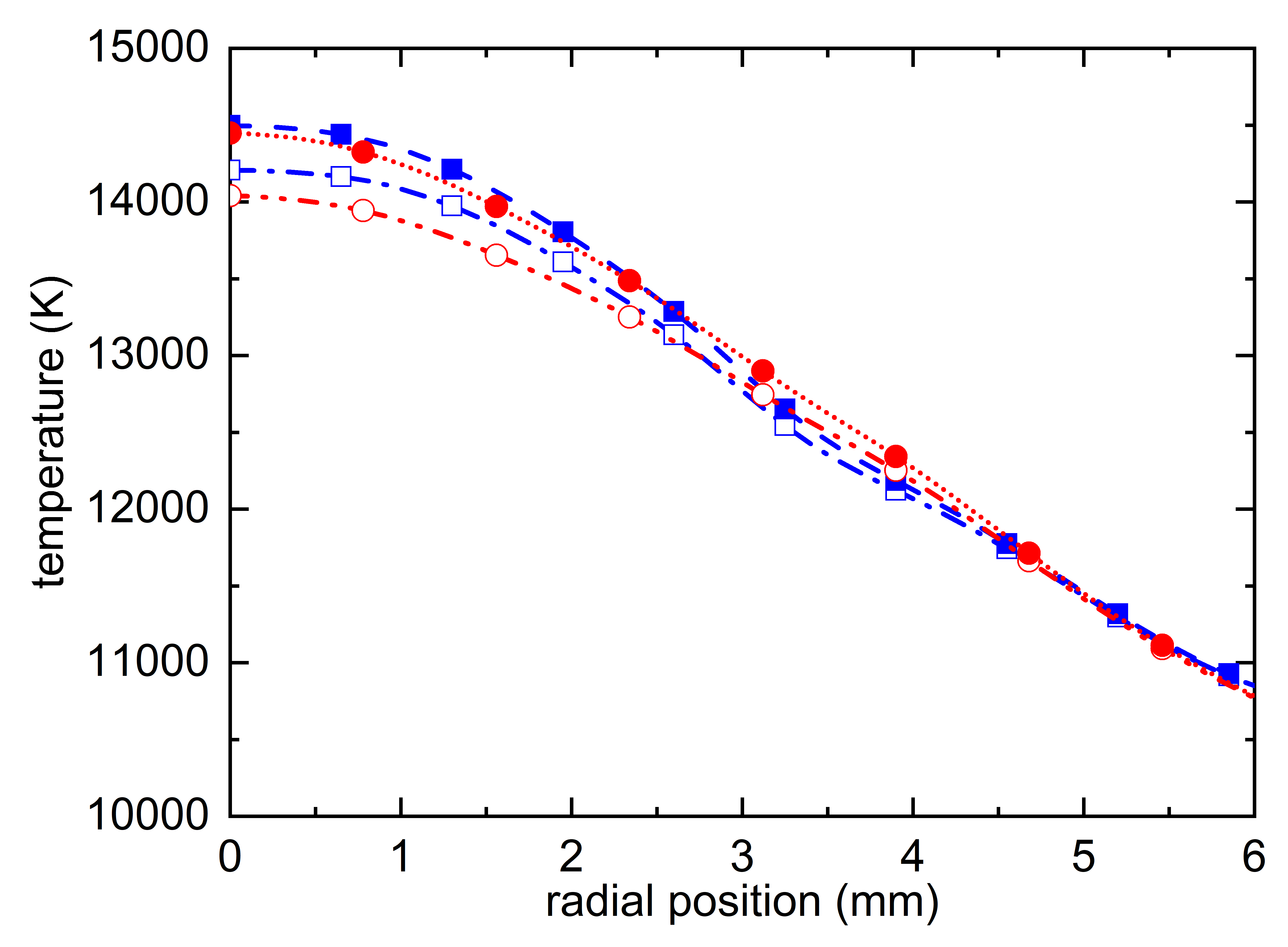}
	\caption{Radial temperature profiles for F\,I at 641\,nm (squares) as well as at 635\,nm (circles) without correction with respect to absorption effects (dashed line) and with iterative absorption correction (solid line), measured 0.6\,ms before current zero at 3.3\,kA instantaneous current.}
	\label{fig:IterResults} 
\end{figure}

Figure\,\ref{fig:IterResults} presents temperature profiles for F\,I at 641\,nm (squares) as well as at 635\,nm (circles) with conditions of the plasma as given in the figures above (10\,kA effective maximum current, 3.3\,kA instantaneous current at the moment of spectrum acquisition, 1.25\,ms before current zero). The lines with open symbols give the temperatures for the zeroth iteration cycle that is the procedure of the calibrated line method (described above) with the assumption of an optically thin plasma. The lines with full symbols give the final temperature profile after iteration. As can be seen from Fig.\,\ref{fig:IterResults} and from Tab.\,\ref{tab:IterResults} the optical depth for the spectral line F\,I at 635\,nm is higher than for F\,I at 641\,nm. After iteration axis temperatures are very similar for both lines. The iteration has obviously a stronger effect on that line with higher absorption F\,I 635\,nm. The temperature increase after iteration is about 300 -- 400\,K.  The temperature correction due to absorption effects is maximum in the arc centre and decreases for increasing radial position, because the optical path length tends towards zero for geometrical reasons and because the emission goes down due to decreasing temperatures. The experimental uncertainty typically is assumed to be about $\pm5\,\%$ that is $\pm700\,K$. This experimental uncertainty mainly arises from systematic uncertainty of the calibration, from uncertainties in atomic data, pressure measurement and from the ill-posed Abel inversion problem plus some statistical uncertainty due to signal fluctuations. Not considering the absorption effect would increase the experimental uncertainty by 50\,\% which is significant, although the absolute error is comparably small. The radial temperature profile is not plotted for radial positions higher than 6\,mm, because spectral radiance becomes such low that experimental uncertainty increases dramatically due to a poor signal to noise ratio. 
\begin{table}
	\centering
	\caption{Results from iteration procedure according to Fig.\,\ref{fig:IterResults} for two spectral lines of F\,I. $\lambda_0$: Centre wavelength of spectral line, $T^{(0)}$: Axis temperature without iteration, $T^\mathrm{(iter)}$ Axis temperature after iteration, $\tau_0$: Optical depth in arc axis and in line centre}
	\begin{tabular}{llll}
		\hline
		$\lambda_0$ (nm) & $T^{(0)}$ (K) & $T^\mathrm{(iter)}$ (K) & $\tau_0$ \\ \hline
		635 & 14040 & 14450 & 0.401 \\ 
		641 & 14210 & 14500 & 0.276 \\ \hline
	\end{tabular}
	\label{tab:IterResults}
\end{table}

\subsection{SF$_6$ switching arc temperatures in a PTFE nozzle}

Spectroscopic measurements have been analysed for three shots in the model breaker. Each shot has been generated with the same electric parameters and the same contact opening speed. The reproducibility of the shots has been proven by voltage, current and pressure measurements. Spectra have been recorded in each of these shots at different times before current zero. As discussed above F\,I spectral line at 641\,nm is less affected by absorption effects compared to F\,I at 635\,nm. Hence, temperature evaluation has been performed with this line. Iterative absorption correction has been performed anyway.
\begin{figure}[ht]
	\centering
	\includegraphics[width=8cm]{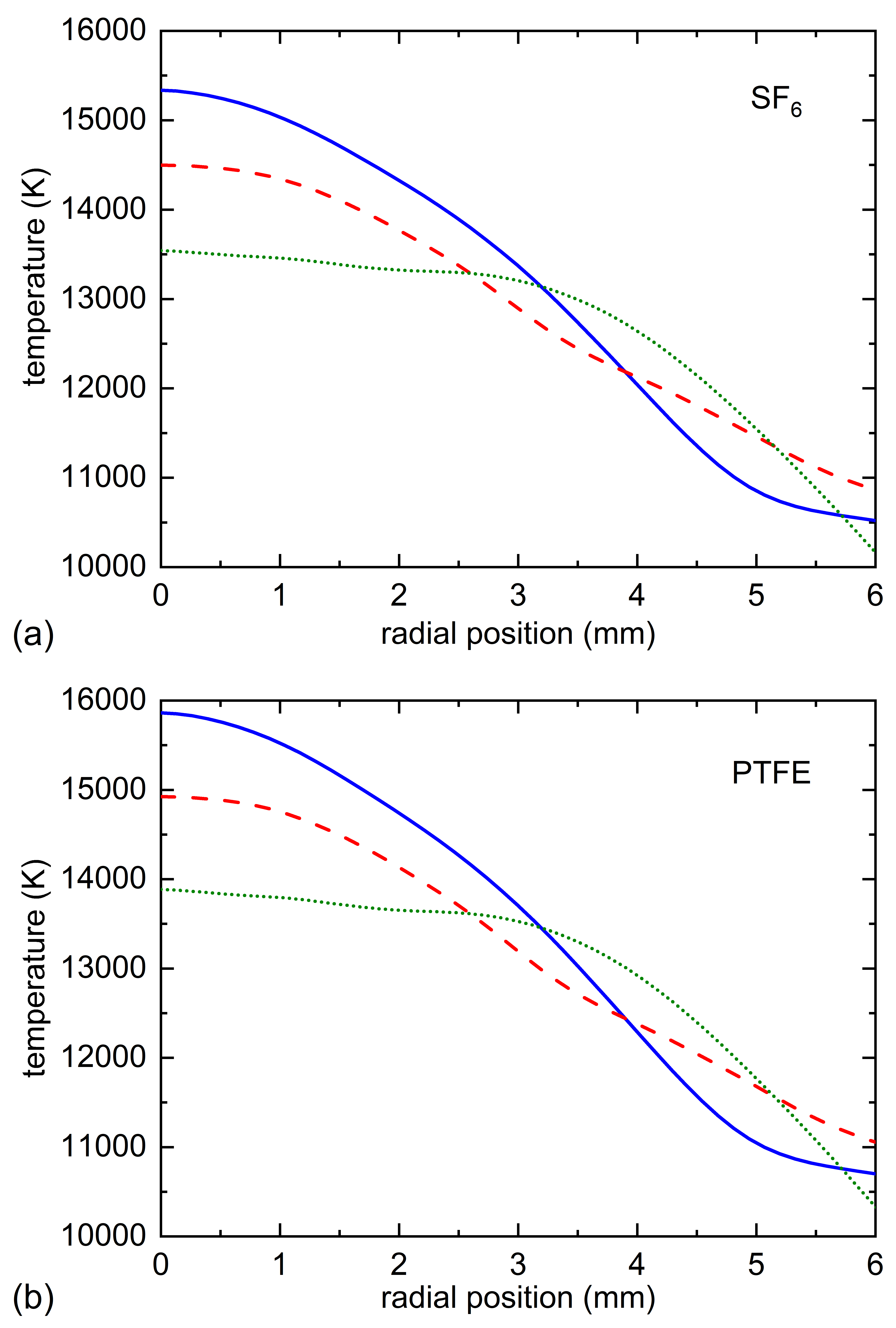}
	\caption{Radial temperature profiles for the three currents at 1.7\,kA (blue solid line), 3.3\,kA (red dashed line), and 3.5\,kA (green dotted line) assuming (a) pure SF$_6$ and (b) pure PTFE plasma composition.}
	\label{fig:SF6PTFE}
\end{figure} 
\begin{table}
	\centering
	\caption{Results from temperature determination according to Fig.\,\ref{fig:SF6PTFE} obtained with F\,I specrral line at 641\,nm for two different plasma compositions, namely pure SF$_6$ and pure PTFE at given pressures $p$. $I$: instantaneous current, $\Delta t$: Time to current zero crossing, $T$: Axis temperature, $\tau_0$: Optical depth in arc axis and in line centre.}
	\begin{tabular}{lllllll}
		\hline
		$I$ (kA) & $\Delta t$ (ms) & $p$ (bar) & $T$@SF$_6$ (K) & $\tau_0$@SF$_6$ & $T$@PTFE (K) & $\tau_0$@PTFE \\ \hline
		1.7 & 0.3 & 6.5 & 15\,340 & 0.284 & 15\,860 & 0.263 \\ 
		3.3 & 0.6 & 6.8 & 14\,500 & 0.276 & 14\,920 & 0.257 \\ 
		3.5 & 0.7 & 6.8 & 13\,540 & 0.195 & 13\,890 & 0.184 \\ \hline
	\end{tabular}
	\label{tab:SF6PTFE}
\end{table}

Figure\,\ref{fig:SF6PTFE} shows radial temperature profiles for three different instantaneous currents after iterative calibrated line method taking account for absorption effects. The temperature profiles are determined assuming a pure SF$_6$ gas in one case and a pure PTFE gas in the other case. 

The relative temperature profile look very similar in both cases which is expected, because the only difference is the plasma composition used for temperature determination. This is verified by the fact that the fluorine density in the semi-logarithmic plot of Fig.\,\ref{fig:dens-eps-theo}(a) representing both considered plasma compositions are just characterized by an offset. The same is true for the derived emission coefficients in Fig.\,\ref{fig:dens-eps-theo}(b), particularly in the temperature range between 10\,000\,K and 16\,000\,K. Therefore, starting the temperature evaluation with the same experimental emission coefficients the final temperature profiles should differ almost by a constant factor. 

The relative contour of temperature profiles is probably more affected by the iterative evaluation depending on the optical depth. In Fig.\,\ref{fig:IterResults} it was already shown, that iteration increases the axis temperature the most whereas absorption effects are less at outer radial positions. However, the overall relative contour of temperature profiles is only slightly modified compared to temperature profiles obtained without iteration\cite{Methling2015p163}. At least, the effect of iteration is less than the effect of current decrease on the radial temperature profile.

Obviously, the absolute temperatures differ significantly assuming SF$_6$ gas or PTFE gas. Table\,\ref{tab:SF6PTFE} summarizes axis temperatures for all considered cases. A lower pre-calculated emission coefficient for the case of PTFE compared to SF$_6$ leads to increased temperatures from calibrated line method. Hence, PTFE temperatures are higher than SF$_6$ temperatures in the arc axis by 350\, up to 500\,K. But absorption effects are less in a PTFE plasma compared to a SF$_6$ plasma because of lower fluorine density in case of PTFE which can be verified by the optical depth $\tau_0$ in the arc axis and in the line centre also given in Table\,\ref{tab:SF6PTFE}. Absorption effects are strongest for plasmas with highest emission in the arc axis that is at 1.7\,kA instantaneous current according to Kirchhoff's law given in Eqn.\,(\ref{eqn:Kirchhoff}). The optical depths range for the conditions studied here between 0.18 and 0.28, which corresponds to an absorption between 16\% and 24\%, respectively. It should be noted that absorption effects could be stronger choosing another spectral line like F\,I at 635\,nm.

The absolute temperature increase due the iterative procedure for the cases presented in Fig.\,\ref{fig:SF6PTFE} is between 190\,K and 370\,K which is between 1.5\,\% and 2.5\,\% of the maximum temperature in the arc axis, respectively. 

As can be seen from Fig.\,\ref{fig:SF6PTFE} the temperature profiles constrict for decreasing instantaneous currents towards current zero. This constriction follows the trend of arc constriction observed in high-speed videos recording the overall intensity distribution in the arc. At the same time the axis temperature increases with decreasing current. This temperature increase probably compensates partly the overall loss of electrical conductivity by the arc constriction. This effect is known from the transition of wall constricted arcs to free burning arcs\cite{Gueye2017p073302,Reichert2015a}, usually induced by a cooling of outer arc layers. In those cases there can exist a transient phase of arc constriction with increased axis temperature. Here, it is expected that a delayed release of PTFE wall material or an inflow of comparably cold SF$_6$ gas during this phase of switch gear operation contribute to a change of arc mode from wall stabilized to free burning arc followed by an arc constriction and an increase of axis temperature. However, with decreasing ohmic heating towards current zero it is expected that the effective temperature over the arc cross section decreases as already observed in literature\cite{Airey1975}. It should be noted that temperature profiles presented in Fig.\,\ref{fig:SF6PTFE} are only 300 -- 400\,$\mu$s apart. Therefore it is concluded that the observed effect seems to be reasonable. However, the extent of temperature increase in the arc axis is surprising. 

Let us consider two facts which may counteract on the apparent temperature increase.
First, this temperature increase could partly balanced by a change of plasma composition. However, the question, which plasma composition does lead to a more realistic temperature profile, cannot be answered for certain. The arc starts in a SF$_6$ gas. As the arc temperature increases within the sine half-wave towards current maximum more and more PTFE is ablated and a dominance of PTFE in the gas phase could be expected. Towards current zero self-blast circuit breakers blow exhaust gas of the nozzle into the arc again. This should be a mixture of SF$_6$ as well as PTFE. This is supported by the fact that the maximum pressure is not higher than 7.5\,bar, which is 1.5\,bar above the chamber base pressure filled with 6\,bar of SF$_6$. Therefore temperatures obtained for plasma compositions of SF$_6$ and PTFE quantify just another contribution to the experimental uncertainty of temperature determination in this study. However, let us assume that for 3.5\,kA there is a pure PTFE plasma and for 1.7\,kA a pure SF$_6$ plasma the temperature increase reduces from approximately 2000\,K to 1500\,K. This is not enough to reverse the temperature increase.

Second, a significant density of fume particles may give rise for an underestimation of axis temperature at elevated currents. However, no abnormally high scattering rate of arc radiation in dark plasma layers close to the nozzle wall has been observed. Even an additional change of transmission by 50\,\% would just lead to a reduction of temperature by 500\,K and therefore could not reverse the temperature trend. It must be emphasized that there is no experimental evidence for such a strong effect of fumes on the spectroscopic results. It can be summarized that not even both discussed facts together are able to reverse the temperature trend but probably may reduce the temperature difference from 2000\,K to 1000\,K in maximum.

Non-LTE effects are not expected to play any role at these pressures and current densities. CFD simulations are appropriate tools to provide deeper insights into the microphysical processes and in parts already confirm the obtained results\cite{Petchanka2015p63,Reichert2015a}.

\section{Summary and outlook}

High-resolution spectra have been measured for an ablation controlled switching arc in a model circuit breaker filled with 6\,bar SF$_6$ and equipped with a PTFE nozzle. The focus was on a switching-off process with 14\,kA peak current and here particularly on its plasma properties close to current zero.

It can be summarized that a Lorentz-like spectral line shape is not a safe proof for optically thin conditions. However, if spectra are calibrated in units of spectral radiance and the radiator density is available, e.g. from pressure and plasma composition, there is an easy way to estimate optical depth of the plasma without any additional experimental effort. Moreover the evaluation of spectral lines according to the calibrated line method under the assumption of optically thin plasma can be modified to allow for the consideration of absorption effects. The iterative method for optical emission spectroscopy making use of Kirchoff's law has been introduced and exemplified by the evaluation of spectral lines of atomic fluorine. It should be mentioned, that this method is applicable to numerous spectroscopic studies in plasmas with unknown optical depth, where absorption effects cannot be excluded.

The plasma investigated here revealed to be neither optically thin nor (extremely) optically thick. Optical depth in the line centre and for the longest optical path through the arc axis $\tau_0$ was found to range between 0.2 and 0.4. Temperature profiles have been corrected taking account of absorption effects, which lead to a temperature increase between 300 and 400\,K at a maximum arc temperature of around 16\,000\,K. This correction is in the order of magnitude of other evaluation uncertainties introduced by absolute calibration, transition probabilities, pressure measurement, and plasma composition. All those uncertainties are mainly of systematic nature. Hence, it is strongly suggested to account for absorption effects, bringing the investigator into the position to decide whether different systematic uncertainties add with the same sign or affect the results in opposite directions. In this study it was found that beside the re-absorption of radiation the addition of PTFE to the plasma composition gives rise to another increase of axis temperature by 350 to 500\,K, which amounts to a total temperature increase of about 700\,K. (see Tab.\,\ref{tab:IterResults} and Tab.\,\ref{fig:SF6PTFE} for F\,I at 641\,nm and 3.3\,kA).

The radial temperature profiles obtained for three different currents less than one millisecond before current zero show a constriction of the arc temperature towards current zero. At the same time axis temperature tends to increase from 14\,000\,K to 16\,000\,K in case of a pure PTFE arc. For pure SF$_6$ temperatures are lower.

The results are valuable for CFD simulation of realistic circuit breakers processes and will help to verify corresponding codes. Although the recovery of the switching path is of particular importance for commercial circuit breakers, temperature measurements during the maximum of a high-current arc would be helpful for the verification of CFD codes. However, the investigations revealed optically thick conditions close to current zero and much higher optical depths are expected in current maximum. Together with a strong continuum emission observed in first preliminary measurements it is obvious that full radiative transfer calculations would be required to obtain deeper insights into the plasma processes and spatial temperature distributions which might be subject to future investigations.

\section*{References}

\bibliographystyle{unsrt}
\bibliography{references} 

\begin{thebibliography}{10}

\bibitem{God00}
D.~Godin, J.~Y. Trepanier, M.~Reggio, X.~D. Zhang, and R.~Camarero.
\newblock {Modelling and simulation of nozzle ablation in high-voltage
  circuit-breakers}.
\newblock {\em J. Phys. D Appl. Phys.}, 33(20):2583--2590, 2000.

\bibitem{Wie91}
W.~L. Wiese.
\newblock {Spectroscopic Diagnostics of Low-Temperature Plasmas - Techniques
  and Required Data}.
\newblock {\em Spectrochim. Acta B}, 46(6-7):831--841, 1991.

\bibitem{Ruch86}
C.~B. Ruchti and L.~Niemeyer.
\newblock {Ablation Controlled Arcs}.
\newblock {\em IEEE Trans. Plasma Sci.}, 14(4):423--434, 1986.

\bibitem{Koz07}
R.~Kozakov, M.~Kettlitz, K.~D. Weltmann, A.~Steffens, and C.~M. Franck.
\newblock {Temperature profiles of an ablation controlled arc in PTFE: I.
  Spectroscopic measurements}.
\newblock {\em J. Phys. D Appl. Phys.}, 40(8):2499--2506, 2007.

\bibitem{Tan97}
Y.~Tanaka, Y.~Yokomizu, T.~Matsumura, and Y.~Kito.
\newblock {The opening process of thermal plasma contacts in a post-arc channel
  after current zero in a flat-type SF6 gas-blast quenching chamber}.
\newblock {\em J. Phys. D Appl. Phys.}, 30(3):407--416, 1997.

\bibitem{Hart98}
K.~T. Hartinger, L.~Pierre, and C.~Cahen.
\newblock {Combination of emission spectroscopy and fast imagery to
  characterize high-voltage SF$_6$ circuit breakers}.
\newblock {\em J. Phys. D Appl. Phys.}, 31(19):2566--2576, 1998.

\bibitem{Eich12}
D.~Eichhoff, A.~Kurz, R.~Kozakov, G.~Gott, D.~Uhrlandt, and A.~Schnettler.
\newblock {Study of an ablation-dominated arc in a model circuit breaker}.
\newblock {\em J. Phys. D Appl. Phys.}, 45(30), 2012.

\bibitem{Methling2015p163}
R.~Methling, St. Franke, D.~Uhrlandt, S.~Gorchakov, F.~Reichert, and
  A.~Petchanka.
\newblock {Spectroscopic Study of Arc Temperature Profiles of a Switching-off
  Process in a Model Chamber}.
\newblock {\em Plasma Physics and Technology}, 2(2):163--166, 2015.

\bibitem{NIST2020}
A.~Kramida, Yu. Ralchenko, J.~Reader, and NIST~ASD Team.
\newblock {NIST Atomic Spectra Database (version 5.7.1)}.
\newblock {[Online]. Available: http://physics.nist.gov/asd [retrieved January
  2020]}, National Institute of Standards and Technology, Gaithersburg,
  Maryland, 2019.

\bibitem{Ebeling1973a}
W.~Ebeling and R.~Sändig.
\newblock {Theory of the lonization Equilibrium in Dense Plasmas}.
\newblock {\em Annalen der Physik}, 483(4):289--305, 1973.

\bibitem{Kahlbaum1990p753}
T.~Kahlbaum and A.~Förster.
\newblock {Thermodynamic properties of nonideal plasmas with multiple
  ionization and Coulomb and hard-core interactions}.
\newblock {\em Laser and Particle Beams}, 8(4):753--762, 1990.

\bibitem{Lochte-Holtgreven1995a}
W.~Lochte-Holtgreven.
\newblock {Evaluation of Plasma Parameters}.
\newblock In W.~Lochte-Holtgreven, editor, {\em Plasma Diagnostics}, page 135.
  American Institute of Physics, New York, 1995.

\bibitem{Zwicker1995a}
H.~Zwicker.
\newblock {Evaluation of Plasma Parameters in Optically Thick Plasmas}.
\newblock In W.~Lochte-Holtgreven, editor, {\em Plasma Diagnostics}, page 214.
  American Institute of Physics, New York, 1995.

\bibitem{Franke2014a}
St. Franke, R.~Methling, D.~Uhrlandt, R.~Bianchetti, R.~Gati, and M.~Schwinne.
\newblock {Temperature determination in copper-dominated free-burning arcs}.
\newblock {\em J. Phys. D Appl. Phys.}, 47(1):015202, 2014.

\bibitem{Petchanka2015p63}
A.~Petchanka, F.~Reichert, R.~Methling, and St. Franke.
\newblock {CFD Arc Simulation of a Switching-off Process in a Model Chamber}.
\newblock {\em Plasma Physics and Technology}, 2(1):63--66, 2015.

\end{thebibliography}

\end{document}